\documentstyle[twocolumn,prd,aps]{revtex}
\begin{document}
\draft
\title{Quark-Gluon Plasma as a Condensate of $Z(3)$ Wilson Lines}
\author{Robert D. Pisarski}
\bigskip
\address{
Department of Physics, Brookhaven National Laboratory,
Upton, New York 11973-5000, USA\\
pisarski@bnl.gov\\
}
\date{\today}
\maketitle
\begin{abstract}
Effective theories for the thermal Wilson line
are constructed in an $SU(N)$ gauge theory at nonzero temperature.
I propose that the order of the deconfining phase transition
for $Z(N)$ Wilson lines is governed by the behavior of $SU(N)$ Wilson lines.
In a mean field theory, the free energy in the deconfined phase is controlled
by the condensate for $Z(N)$ Wilson lines.  
Numerical simulations on the lattice, 
and the mean field theory for $Z(3)$ Wilson lines, 
suggest that about any finite temperature transition in QCD, 
the dominant correlation length 
increases by a large, uniform factor, of order five.
\end{abstract}

A new phase of matter, the Quark-Gluon Plasma, might be produced
in the collisions of large nuclei at very high
energies.  By asymptotic freedom, the pressure approaches the
ideal gas value in the limit of high temperature, and so it is
natural to think of the high temperature phase of QCD as
a gas of quasiparticles \cite{quasi,htl}.

It is known, however, that the {\it high} temperature phase of
a purely glue theory is like the {\it low} temperature phase of a 
spin system.  The magnetization in the high temperature
phase of a $SU(N)$ gauge theory is a $Z(N)$ spin,
proportional to the trace of the thermal Wilson line \cite{svet}.

In this paper I construct effective lagrangians for the thermal
Wilson line, considered as a full $SU(N)$ matrix,
as well as its trace.  This leads to novel sigma models
of adjoint $SU(N)$ fields.  Although the critical behavior is
inexorably governed by the fixed point of $Z(N)$ spins \cite{svet}, the
$SU(N)$ spins can be important. In particular, they 
help explain why the order of the deconfining 
transition appears to change with $N$: from
second order for $N=2$ \cite{twoold,two,kiskis_vranas}, 
to weakly first order for $N=3$ \cite{three,freeenergy,potentials,gavai},
to first order for $N \geq 4$ \cite{rpfour,ohta_wingate}.  
Further, the picture of the high temperature phase is turned
on its head: the pressure isn't due to quasiparticles \cite{quasi,htl},
but is a potential for a condensate of $Z(N)$ Wilson lines.
A mean field theory then suggests that 
because the deconfining transition in pure glue $SU(3)$ is weakly
first order, QCD is 
near a critical point.  About the transition, the dominant
correlation lengths
increase by a large factor, of order five \cite{potentials}.

I concentrate on the pure glue theory; later I argue why this is 
legitimate, using the lattice data and the effective theory.
The thermal Wilson line is \cite{svet,gpy}
\begin{equation}
{\bf L}(x) = {\cal P} \exp \left( i g \int^{1/T}_0 A_0(x,\tau) \, 
d\tau \right) \; ,
\label{e1}
\end{equation}
where $\cal P$ is path ordering, $g$ is the gauge coupling constant,
$A_0$ is the time component of the vector potential
in the fundamental representation, 
$x$ is the coordinate for three spatial 
dimensions, and $\tau$ that for euclidean time at a temperature $T$.
The Wilson line in (\ref{e1}) is a product of $SU(N)$ matrices,
and so is itself a $SU(N)$ matrix, satisfying
\begin{equation}
{\bf L}^\dagger(x) \, {\bf L}(x) = {\bf 1}\;\;\; ,\;\;\;
\det({\bf L}(x) ) = 1 
\label{e2}
\end{equation}
Without quarks, the allowed gauge transformations are periodic up
to an element of a global $Z(N)$ symmetry \cite{svet}:
\begin{equation}
{\bf L}(x) \rightarrow \exp(2 \pi i/N) \; \Omega^\dagger(x)
{\bf L}(x) \Omega(x) \; ,
\label{e3}
\end{equation}
$\Omega(x)=\Omega(x,0)$.  ${\bf L}(x)$ transforms as
an adjoint field under local
$SU(N)$ gauge transformations in three dimensions, 
and as a vector under global $Z(N)$ transformations.

Effective theories for ${\bf L}(x)$ are dictated by 
the symmetries of (\ref{e3}).  I begin with the nonlinear form,
where (\ref{e2}) are taken as constraints on $\bf L$.  
I construct an effective theory in three spatial dimensions, 
valid for distances $\gg 1/T$, by coupling
the gauge potentials for static magnetic fields, 
the $A_i(x)$'s, to the Wilson line, ${\bf L}(x)$:
\begin{equation}
{\cal L}_0 \; = \;
\frac{1}{2} {\rm tr} \left( G^2_{ij} \right)
+ T^2 a_1 \, {\rm tr} |D_i {\bf L}|^2 \;\;\; , \;\;\; 
a_1 = \frac{1}{g^2} + \ldots
\label{e4}
\end{equation}
The first term is the standard lagrangian for 
static $A_i$ fields (by choice, $A_i$ has dimensions of mass;
all lagrangians have dimensions of (mass)$^4$).
In the second term, I start with
electric part of the gauge lagrangian, $\sim {\rm tr}|D_i A_0|^2$,
and assume that it transmutes into a gauge invariant
kinetic term for ${\bf L}(x)$.
This is the continuum form of the lattice model of
Banks and Ukawa \cite{banks_ukawa}.

Notice the 
factor of $T^2$ in front of the kinetic term for ${\bf L}$.
This arises because the Wilson line is a phase in
color space, and so every element is a dimensionless
pure number.  Thus in any effective lagrangian, dimensions can only be
made up by powers of the temperature $T$.  

Consider the somewhat peculiar
limit in which one drops the coupling to the $A_i$'s,
by taking $g\rightarrow 0$, but retains ${\bf L} \neq {\bf 1}$.
Then (\ref{e4}) reduces a nonlinear sigma model in three dimensions,
with lagrangian $\sim {\rm tr} |\partial_i {\bf L}|^2$.
With the constraints of (\ref{e2}), 
the theory is invariant under ${\bf L} \rightarrow 
\Omega_1 {\bf L} \Omega_2$, where $\Omega_{1}$ and $\Omega_2$ are independent,
constant $SU(N)$ matrices.
This is an enhanced global symmetry 
of $SU(N) \times SU(N)$ (times the usual global $Z(N)$ symmetry).
As a sigma model, it is possible to impose other 
constraints upon $\bf L$ beyond those of (\ref{e2}).  For example,
requiring ${\rm tr}{\bf L}$ to be some fixed
number produces a sigma model on a symmetric space \cite{symmetric}.

At nonzero coupling, $\bf L$ is simply an adjoint field under the
local $SU(N)$ symmetry.
Even with the constraints of (\ref{e2}), the reduced
symmetry implies that {\it many} more terms arise:
instead of ${\rm tr}{\bf L}$ being fixed, as for a symmetric space,
arbitrary traces, such as ${\rm tr}{\bf L}^p$ for
integer $p$, are allowed.
Mathematically, ${\rm tr} {\bf L}^p$ is related to the trace
of the Wilson line in higher representations \cite{drouffe_zuber}.

At one loop order, the terms up to fourth order in
$A_0$ have been computed.  The quadratic term, 
$\sim {\rm tr}(A^2_0)$ \cite{debye,laine_philipsen}, 
is the Debye mass for the gluon.
For $N \geq 4$, there are two independent quartic terms, 
$\sim ({\rm tr}(A^2_0))^2$ 
and $\sim {\rm tr}(A^4_0)$ \cite{quartic,kajantie}, which 
represent a potential for $A_0$.  From (\ref{e1}), it
is easy to turn a potential for $A_0$ into one for $\bf L$:
\begin{equation}
{\cal L}_1 = T^4 \left(
c_2 |{\rm tr} {\bf L}|^2 + c_4 |{\rm tr} {\bf L}^2|^2
+ c_4' \left( |{\rm tr} {\bf L}|^2 \right)^2 \right) \; .
\label{e5}
\end{equation}
Expanding to $\sim A_0^4$ fixes
$c_2 = -(4 + 3/\pi^2)/9$,
$c_4 = + ( 1 + 3/\pi^2 )/36$, and $c_4' = 0$.
The rational terms in $c_2$ and $c_4$ are from the Debye mass,
while those $\sim 1/\pi^2$ arise from quartic
terms in the potential for $A_0$.  Only two constants,
$c_2$ and $c_4$, are needed to fit three terms in the $A_0$ potential.
If $N_f$ flavors of massless quarks are included,
$c_2$ and $c_4$ change, while $c_4'$ is then nonzero.

The signs of $c_2$ and $c_4 $ are interesting.
As ${\bf L} \sim - g^2 A_0^2$, a positive Debye mass corresponds
to negative $c_2$.  The coupling $c_4$ is like
the quartic term in $A_0$, and so positive.
Negative $c_2$ favors condensation in a direction in
which $|{\rm tr}{\bf L}|^2$ is maximized.  This happens when
$\bf L$ is an element of the center \cite{wingate}, 
\begin{equation}
\langle {\bf L} \rangle = \exp(2 \pi i j/N) \; \ell_0\; {\bf 1} \; ,
\label{e7}
\end{equation}
$j = 0...(N-1)$.  Different $j$ are the usual $N$ degenerate 
vacua of the broken $Z(N)$ global symmetry.  

In (\ref{e7}) I introduce an expectation value, 
$\ell_0 = \langle \ell \rangle$, where $\ell$ is defined in (\ref{e10}).
In perturbation theory, $\ell_0 = 1$,
but $\ell_0$ is a function of temperature;
it vanishes at the critical temperature,
$T_c$, and in the confined phase, for $T < T_c$.

I assume that in the deconfined phase, $T > T_c$,
the stable vacuum is that which maximizes $|{\rm tr}{\bf L}|^2$, so 
that $\bf L$ condenses as in (\ref{e7}).
An expectation value for a field in the fundamental
representation always breaks the gauge symmetry, but uniquely
for an adjoint field, a vacuum expectation value 
proportional to the unit matrix 
does {\it not}: (\ref{e7}) is invariant under
arbitrary local gauge rotations.  Similarly, 
the adjoint covariant derivative in (\ref{e4}) is $D_i {\bf L} =
\partial_i {\bf L} - i g [A_i,{\bf L}]$, so with (\ref{e7}), the
static magnetic gluons do not acquire a mass
when $\bf L$ condenses, $\ell_0 \neq 0$.  Thus (\ref{e7}) is 
the nonperturbative statement that electric screening does not 
generate screening for static magnetic 
fields \cite{gpy,kajantie,negative}.

The terms in (\ref{e5}) are invariant under a global symmetry of
$U(1)$.  There are also terms which reduce
this $U(1)$ to $Z(N)$.  For $N=3$, the simplest examples include
\begin{equation}
\det{\bf L} + {\rm c.c.} \; ,\;
({\rm tr} {\bf L})^3 + {\rm c.c.} \; ,\;
{\rm tr} {\bf L}\; ({\rm tr} {\bf L}^2) + {\rm c.c.} \; .
\label{e8}
\end{equation}
The first term, $\det{\bf L}$, is $SU(3) \times SU(3)$ symmetric,
while the others are only $SU(3)$ symmetric.

There are also a wide variety of kinetic terms possible.  These include
$|\partial_i {\rm tr}{\bf L}|^2$,
$|\partial_i {\rm tr}{\bf L}^2|^2$, and
$|{\rm tr}{\bf L}|^2 {\rm tr}|D_i {\bf L}|^2$,
amongst others.  At one loop order, the kinetic term in (\ref{e3})
is renormalized, and terms such as these may be generated;
present calculations cannot distinguish \cite{interface}.
Even for $g=0$, none of these new kinetic terms
are invariant under $SU(N) \times SU(N)$.

The potential in (\ref{e5}) is only illustrative.
In perturbation theory, one expands about ${\bf L} \sim {\bf 1}$,
which does not allow one to uniquely fix the coefficients of
a potential for $\bf L$.  Through numerical
simulations, effective theories for $A_0$ \cite{kajantie},
and those for $\bf L$, could be matched by comparing physical
correlation lengths at an intermediate temperature scale, say
at several times the critical temperature.  

Now consider a point of second order transition, where
$\ell_0(T) \rightarrow 0$.  Then powers of $\bf L$ are suppressed,
and it is sensible to construct a linear sigma model.
This is done by introducing an
``block spin'' ${\bf L}$, formed by a gauge covariant average of
$\bf L$ over some region of space \cite{gauge}.
Any $SU(N)$ matrix can be written as
\begin{equation}
{\bf L}(x) = \ell(x) {\bf 1} + 2 i  \tilde{\ell}_a(x) t^a \; ,
\label{e10}
\end{equation}
where $t^a$ are the generators of $SU(N)$, $a = 1\ldots(N^2-1)$.  
For general $N$,
$\ell$ and $\tilde{\ell}_a$ are complex valued, and
(\ref{e2}) imposes $N^2 + 1$ constraints.

I start with the case of two colors, which is special.
Four constraints of (\ref{e2}) are satisfied
in an especially simple manner: the imaginary parts of $\ell$ and
$\tilde{\ell}_a$ vanish.  This leaves one constraint,
which is $\ell^2 + \tilde{\ell}_a^2 = 1$; thus
$\ell$ and $\tilde{\ell}_a$ form a vector representation of
$SU(2) \times SU(2) = O(4)$.  
After averaging, the constraint on the $O(4)$ norm is lost,
as is typical in a linear model.
Averaging still leaves 
$\ell$ and $\tilde{\ell}_a$ as real valued fields, though.
Up to quartic order, the most general lagrangian is
\begin{equation}
{\cal L} \; = \;
\frac{1}{2} {\rm tr} \left( G^2_{ij} \right)
\; + \; \frac{1}{2} (\partial_i \ell)^2 \; + \; 
{\rm tr}|D_i \tilde{\ell}|^2  \;
\label{e11}
\end{equation}
$$
- m_1( \ell^2 + \tilde{\ell}_a^2 ) 
- m_2 \ell^2 
+ \lambda_1 ( \ell^2 + \tilde{\ell}_a^2 )^2 
+ \lambda_2 \, \ell^4 + \lambda_3 \, \ell^2 \, \tilde{\ell}_a^2  \; .
$$
The $\ell$-field is a color singlet, while $\tilde{\ell}_a$ is an
adjoint $SU(2)$ field.
The terms $\sim m_1$ and $\lambda_1$ are $O(4)$ symmetric;
with the kinetic terms, they correspond to the gauged nonlinear 
sigma model of (\ref{e4}).  The other terms, $\sim m_2$,
$\lambda_2$ and $\lambda_3$, correspond to the
potential for the Wilson line in (\ref{e5}).  A factor of $T$
has been absorbed into the definition of $\ell$ and $\tilde{\ell}_a$.

When $N \geq 3$, the constraints of (\ref{e2}) are nonlinear.
Since a sum of two special unitary matrices is not necessarily special
unitary, the average $\bf L$ must be taken to be a complex
$N \times N$ matrix.  Thus $\bf L$ includes 
a complex valued, color singlet field, $\ell$, which I call a $Z(N)$ spin,
and  a complex valued, color adjoint field, $\tilde{\ell}_a$, which
I call a $SU(N)$ spin.

Linear models like (\ref{e11}) can be written down for $N \geq 3$, 
although there is a plethora of terms.  At quartic order there is 
one term which is $O(2 N^2)$ symmetric, 
$({\rm tr} {\bf L}^\dagger {\bf L})^2$,
another which is $SU(N) \times SU(N)$ symmetric, 
${\rm tr} ({\bf L}^\dagger {\bf L})^2$,
and terms which are only invariant under $SU(N)$, such as
($\tilde{\ell} \equiv \tilde{\ell}_a t_a$)
$$
(|\ell|^2)^2 \; , \;
\ell \, {\rm tr}(\tilde{\ell}^\dagger)^2 \ell + {\rm c.c.} \; , \;
\ell^2 \, {\rm tr}(\tilde{\ell}^\dagger)^2 + {\rm c.c.} \; , \;
$$
\begin{equation}
({\rm tr} \tilde{\ell}^\dagger \tilde{\ell})^2 \; , \;
|{\rm tr} \tilde{\ell}^2|^2 \; , \;
{\rm tr} (\tilde{\ell}^\dagger \tilde{\ell})^2 \; , \;
{\rm tr} (\tilde{\ell}^\dagger)^2 \tilde{\ell}^2 \; .
\end{equation}

These models give a qualitative picture
of the deconfining phase transition:
I assume that while only the $Z(N)$ $\ell$-spins condense,
$\langle \ell \rangle \equiv \ell_0 \neq 0$,
that the transition is driven
by the behavior of the $SU(N)$ $\tilde{\ell}_a$-spins.  
This picture is based on the nonlinear model: at weak coupling,
(\ref{e4}) dominates other terms, such as (\ref{e5}), by $\sim 1/g^2$.
Now certainly all coupling constants change with $T$, 
as can be seen from the temperature
dependence of the Debye mass \cite{laine_philipsen}.  Nevertheless, 
I assume that the $SU(N)$ $\tilde{\ell}_a$-spins dominate
right down to the point of the deconfining phase transition.
The only purpose of terms such as (\ref{e5}) is to ensure that
condensation which respects the local $SU(N)$ symmetry, (\ref{e7}),
is favored.

For two colors, the influence of the $SU(2)$ $\tilde{\ell}_a$-spins
on the $Z(2)$ $\ell$-spins is subtle.  Assume that only 
the $O(4)$ symmetric mass, $m_1$, changes.
The phase transition in a gauged $SU(2)$ model is known
from lattice studies of the electroweak phase transition \cite{rp_crit}.
I assume that one is always in an extreme ``type-II'' regime, so that
the second order $O(4)$ transition of the model with $g=0$
(the point $B_2$ of fig. (1) in \cite{rp_crit}) is washed
out by confinement of nonabelian gauge fields.  
The only transition is 
a point at which the $Z(2)$ $\ell$-spins become massless; the 
$SU(2)$ $\tilde{\ell}_a$-spins 
are always massive.  Lattice studies confirm a second order transition
in the $Z(2)$ universality class \cite{two}.  

For $N \geq 3$, the $SU(N)$ $\tilde{\ell}_a$-spins 
can have first order transitions.
This is because in the absence of
gauge fields, $SU(N) \times SU(N)$ spin models have
first order transitions both
in mean field theory \cite{meanfield} and
in an expansion about $4-\epsilon$ dimensions
\cite{rp_crit}.  As suggested in \cite{rp_crit},
in the extreme type-II regime, confinement of the
gauge fields need not wash out the first
order transition of $SU(N)\times SU(N)$ spins
(above the point $B_3$ in fig. (2) of \cite{rp_crit}), 
and so the deconfining transition can remain first order.
In particular, the transition
can be of first order as $N\rightarrow \infty$.
This is in accord with a lattice analysis of Gocksch and Neri 
\cite{gocksch_neri,potts}, and 
contrary to previous speculation \cite{rpfour,largeN}.

For three colors, this implies that the deconfining transition is
of first order not only because of cubic invariants \cite{svet}, 
as in (\ref{e8}), but because of the dynamics of $SU(3)$ 
$\tilde{\ell}_a$-spins.  Relative to the ideal gas, 
the latent heat for three colors is 
$\sim 1/3$ \cite{three}.  
As could have been guessed from the lattice data alone, perhaps the
deconfining transition is 
weakly first order for $N=3$ because it is near the second order transition
for $N=2$.  Thus it is of value to know how the latent 
heat for $N=4$ compares to that for $N=3$: is it 
more strongly first order,  
such as $\sim (N-2)/N$ as $N \rightarrow 2$, 
or more weakly first order, like $\sim 1/N$ as $N \rightarrow \infty$
\cite{rpfour}?  

Whatever the order of the deconfining phase transition, one can write
a mean field theory in which the free energy in the deconfined
phase is controlled by a potential for the $Z(N)$ Wilson lines.
For three colors, this is \cite{svet,rpfour}:
\begin{equation}
{\cal V} \; = \;
\left( - 2 b_2 \, |\ell|^2
+ b_3 ( \ell^3 + (\ell^*)^3 )
+ (|\ell|^2)^2 \right) \; b_4 \; T^4 \; .
\label{e12}
\end{equation}
$\ell$ is complex valued, so when $b_3 \neq 0$, the global
symmetry is reduced 
from $O(2)$ to $Z(3)$.  The coupling $b_3$ must be small for the 
transition to be weakly first order \cite{mass}, so for now I ignore it,
considering the potential just as a function of $b_2$ and $b_4$.
This is similar to the case of two colors, 
where $\ell$ is a real field, and the potential is just a sum
of two terms, $\sim b_2 \ell^2$ and $\sim b_4 \ell^4$ \cite{critical}.

In speaking of the Wilson line, implicitly I assume that
it is possible to extract a renormalized value \cite{trace}
from the bare quantity \cite{bare}.  
If so, then given $\ell_0(T)$ and the pressure, one could fit
to a potential like (\ref{e12}); for example, is it necessary
to include higher powers of $\ell$ in $\cal V$?

The novel aspect of (\ref{e12}) is my insistence that 
because $\ell$ is a {\it dimensionless} field, the
dimensions in $\cal V$ must be made up by the temperature, $T$.
In mean field theory, $b_4$ is taken as constant, and
$b_2$ varies with temperature, vanishing at $T_c$.
The pressure is given by the minimum of the potential,
$p = b_2^2 b_4 T^4$, and vanishes in the confined phase, $T < T_c$.
That is, with the overall $T^4$ in the potential, the pressure
is like a gas of quasiparticles, albeit with a variable number of
degrees of freedom, which vanish at $T_c$.

At high temperature, 
$b_2 \rightarrow 1$ so that $\ell_0 \rightarrow 1$.
The quartic coupling is fixed by the ideal gas limit:
if $n_\infty  = p/T^4$ as $T \rightarrow \infty$, $b_4 = n_\infty$.
Lattice simulations 
\cite{twoold,two,kiskis_vranas,three,freeenergy,potentials,gavai}
find that the $p/T^4$ is relatively flat down to a scale which is
several times $T_c$, call it $\kappa T_c$; the 
same is found from resummations of perturbation theory \cite{htl}.  
($\kappa$ might
be defined as the lowest value where $\ell_0 \approx 1$.)
Hence I assume that $b_2$ and
$b_4$ are slowly varying down to $\kappa T_c$.  

Between $\kappa T_c$ and $T_c$, I assume that $b_4$ is essentially
constant, while $b_2$ varies.  
In particular, the trace of the energy momentum tensor, divided by $T^4$, is 
$(e - 3p)/T^4 = T \partial(b_2^2 b_4)/\partial T$.
Lattice simulations find that this quantity has a peak
just above $T_c$
\cite{twoold,two,kiskis_vranas,three,freeenergy,potentials,gavai}:
this is then due 
to the rapid variation of $b_2$ with temperature \cite{critical}.

In the $Z(N)$ mean field theory, the pressure includes
only the contribution of the potential, and nothing from 
fluctuations in the effective fields, either from the
$Z(N)$ $\ell$-spins or the $SU(N)$ $\tilde{\ell}_a$-spins.  
Fluctuations in these
fields do, of course, contribute to the pressure at all temperatures.
Since by construction the pressure in the mean
field approximation vanishes for $T < T_c$,  one condition for its
validity is that the pressure in the confined phase is small.
This is what present lattice simulations find.  Physically, these
fields don't contribute much to the pressure because they are heavy:
the $SU(N)$ $\tilde{\ell}_a$-spins always so, and the 
$Z(N)$ $\ell$-spins usually so.  For two or three colors, 
the $\ell$-spins do become light in a 
narrow band in temperature about $T_c$, where 
mean field theory fails \cite{critical}.

Ignoring fluctuations about the mean field theory is also justified
from the viewpoint of an expansion in a large number of colors, 
$N \rightarrow \infty$ \cite{gocksch_neri,largeN,thorn}.  The free
energy in the confined phase is of order one, while it is
$\sim N^2$ in the deconfined phase.  The term $\sim N^2$ in the 
free energy is due {\it entirely} to the condensate, taking $b_2 \sim 1$ and
$b_4 \sim N^2$.  Even though there are
$\sim N^2$ of them, the $SU(N)$ $\tilde{\ell}_a$-spins only contribute
to the pressure at $\sim 1$, since they are bound into
color singlet glueballs.  The $Z(N)$ $\ell$-spins also
contribute $\sim 1$ to the free energy.

I have concentrated on the pure glue theory because numerical
simulations have demonstrated the following remarkable property
\cite{three,freeenergy}.  
If $p/(n_\infty T^4)$ is plotted versus $T/T_c$, the resulting curve
is nearly universal, and looks very similar whether or not there
are dynamical quarks present.
The present model predicts that the pressure is the same
because the (renormalized \cite{trace}) Wilson line is the same.
In terms of the potential, (\ref{e12}),
the differences in the ideal gas values, $n_\infty$, 
are absorbed into $b_4$, with the same $b_2(T/T_c)$.

Quarks act like a background magnetic field for the 
real part of $\ell$ \cite{banks_ukawa,quark}.  Because
the pure glue transition is weakly first order, it is not
difficult for quarks to wipe out the deconfining
transition altogether, leaving either a chiral transition,
or just crossover behavior.  
Even so, what is relevant here is that for 
three colors and two or three flavors of quarks,
the pressure for $T < T_c$ is always much smaller than that for
$T > T_c$; that is, up, down, and strange quarks act like
a {\it weak} magnetic field for the $Z(3)$ $\ell$-spins.

I thus come to the central physical point of this paper.  The lattice
tells us that the deconfining transition in pure glue $SU(3)$
theory is close to the second order transition for $SU(2)$; further,
that the effects of quarks are small, except close to
$T_c$.  I suggest that what is
important is {\it not} whether the {\it weakly} first order transition
persists with quarks, but that the {\it nearly} second order transition
very well might.  In the pure glue theory, as $T \rightarrow T_c^+$
the ratio of the screening mass to the temperature 
decreases by a factor of ten:
from $\sim 2.5$ at $T \sim 2 T_c$, to $\sim .25$
at $T \sim T_c^+$ \cite{potentials}.  
Similarly, the string tension at $T \sim T_c^-$ is 
ten times smaller than that at zero temperature \cite{potentials}.
With quarks, the increase in the correlation length
for $\ell$ is presumably less, maybe not ten,
but perhaps a factor of five or so.  And most importantly, 
if the pressure below $T_c$ is small, 
it might be justified to use the $Z(3)$ mean field theory.

If the chiral order parameter is $\Psi$, then 
it couples to $Z(3)$ $\ell$-spins through the coupling
$
+ |\ell|^2 {\rm tr}\left( \Psi^\dagger \Psi \right) \; .
$
Lattice simulations find that the
chiral and deconfining transitions occur at approximately the
same temperature.  This naturally results if this coupling
constant is positive, as condensation in one field tends to suppress
condensation in the other.  Coherent oscillations in
the $\ell$-field couple to light mesons through
such a term, and can produce large fluctuations in the average
pion momentum \cite{ad}.

This uniform increase in correlation lengths near $T_c$
is a unique prediction of the $Z(3)$ mean field theory.
In quasiparticle models of the quark-gluon plasma,
the pressure is tuned to vanish at $T_c$ by the introduction of
a bag constant.  In order for the energy to decrease
as $T \rightarrow T_c^+$, though, the quasiparticles
must become {\it heavier}, not lighter; that is,
instead of increasing, most correlation lengths decrease \cite{quasi,htl}.

At nonzero quark chemical potential, $\mu$, 
presumably there is little change if the quarks are hot and dilute:
for small $\mu/T$, the $Z(3)$ $\ell$-spins should still exhibit
nearly second order behavior.
I contrast this with the (possible) critical
endpoint of the chiral transition in the $\mu-T$ plane \cite{rss}.
The correlation length of the sigma meson truly diverges at
the critical endpoint, but this only occurs at one special value of
$\mu$.  Moreover, the sigma meson does not
dominate the free energy, nor generic particle production.
For cold, dense quark matter, $\mu \gg T$, I do not see why 
$Z(3)$ $\ell$-spins should dominate the free energy.

I conclude by noting that the generalization
of the Debye mass term, $\sim {\rm tr}A_0^2$, to
real scattering processes produces hard
thermal loops \cite{a0}.  This is then the
first term in an {\it infinite} series of such terms, continuing
$\sim {\rm tr}A_0^{4}$, {\it etc.}  
The natural expansion is not in powers of 
$A_0$, but in powers of the Wilson line, as in
(\ref{e5}).  It is then of great interest to know the analytic continuation 
of the Wilson line to real scattering processes \cite{analytic}. 

I benefited from discussions with K. Eskola, F. Gelis,
K. Kajantie, F. Karsch, C. P. Korthals-Altes, M. Laine, D. Miller,
S. Ohta, A. Peshier, D. H. Rischke,
M. Wingate (for the argument in \cite{wingate}), L. Yaffe, 
and, especially, M. Creutz.
This work is supported by DOE grant DE-AC02-76CH00016.


\begin{references}
\bibitem{wingate}
${\bf L}$ can be diagonalized by a global gauge rotation.  
For a $U(N)$ matrix, 
$|{\rm tr}{\bf L}|^2$ is maximized when $\bf L$ is
a constant phase times the unit matrix; for a $SU(N)$ matrix,
then, ${\bf L}$ must be an element of the center.
\bibitem{negative}
For discussion, consider what happens if the Debye mass is
negative, so that the signs of $c_2$ and $c_4$ are
switched, with $c_4 < 0$ and $c_2 > 0$.  In this
case, the favored vacuum maximizes $|{\rm tr}{\bf L}^2|^2$,
with ${\bf L}^2$ an element of the
center; then the vacuum might break $SU(N)$ to $SU(N-1)$.
\bibitem{gauge}
To ensure that the average $SU(N)$ Wilson line transforms gauge covariantly,
as in (\ref{e3}), 
the average is over ``U'' shaped Wilson paths: averaging over $x'$, the
path is spatial from $(x,0)$ to $(x',0)$, temporal up to $(x',1/T)$, 
and finally, spatial back to $(x,1/T)$.
\bibitem{mass} 
$b_3$ is measurable from
the two point function of the Wilson line near $T_c$.
From (\ref{e12}), the mass squared for the real part of $\ell$
is $\sim \ell_0^2/a_1$, while that for
the imaginary part is $\sim b_3 \ell_0/a_1$, with
$a_1$ from (\ref{e4}).  Given $\ell_0$, $b_3$ can
be measured by comparing these two masses.  This should be
done not at high temperature, but close to $T_c$.  At high
temperature, $\bf L$ is nearly an $SU(N)$ matrix and $\ell_0 \approx 1$.
The real and imaginary parts of $\ell$ 
satisfy a geometric constraint, and cannot be varied independently.  
The constraint is invalid near $T_c$, where
the average $\bf L$ is not an $SU(N)$ matrix.  
$\cal V$ is applicable for real $\ell$ (or $Z(3)$ transforms thereof)
at all $T$.
\bibitem{critical}
Mean field theory fails close to a second order transition,
but the effect of nontrivial critical exponents can be included,
as in (12) of \cite{twoold}.
The critical point for two colors is $Z(2)$, while three colors
is near an $O(2)$ fixed point.
\bibitem{analytic}
Define null vectors $K=(i,\hat{k})$ and
$K_*=(-i,\hat{k})$, $\hat{k}^2 = 1$, and
gauge ``potentials'',
$${\cal A}^\alpha = \frac{1}{D\cdot K} K^\mu G^{\mu \alpha} \;\;\; , \;\;\;
{\cal A}^\alpha_* = \frac{1}{D\cdot K_*} K_*^\mu G^{\mu \alpha}.$$  
With $m^2_g = (N + N_f/2) g^2 T^2/3$ the Debye mass squared,
and $\int d\Omega_{\hat{k}}$ the normalized integral over $\hat{k}$,
the usual lagrangian for hard thermal loops is 
${\cal L}_{htl} = m^2_g \int d\Omega_{\hat{k}} {\rm tr} 
({\cal A}_\alpha)^2$ \cite{a0}.  This does not look like 
the expansion of a Wilson line, though.  However, 
an equivalent form is 
$$
{\cal L}_{htl} = m^2_g \int d\Omega_{\hat{k}} 
\, {\rm tr} \left( {\cal A} \cdot K_* \;
{\cal A}_* \cdot K \right) .
$$
which does look like an expansion of exponentials of 
${\cal A} \cdot K_*$ and ${\cal A}_* \cdot K$.
\end{references}
\end{document}